\documentstyle[aps,aipbook,floats]{revtex}
\begin{document}
\title{An Origin for Pulsar Kicks in\\ Supernova Hydrodynamics}
\author{Adam Burrows and John Hayes$^{\dagger}$}
\address{$^{\dagger}$Departments of Physics and Astronomy\\
University of Arizona\\
Tucson, Arizona, USA 85721\\}
\maketitle

\begin{abstract}
It is now believed that pulsars comprise the fastest population of stars in the
galaxy.  With inferred mean, root-mean-square, and maximum 3-D pulsar
speeds of $\sim$300-500 km/s, $\sim$500 km/s, and $\sim$2000 km/s,
respectively,
the question of the origin of such singular proper motions becomes acute.
What mechanism can account for speeds that range from zero to twice the
galactic escape velocity?  We speculate that a major vector component of a
neutron star's
proper motion comes from the hydrodynamic recoil of the nascent neutron star
during
the supernova explosion in which it is born.  Recently, theorists have shown
that
asymmetries and instabilities are a natural aspect of supernova dynamics.  In
this paper,
we highlight two phenomena: 1) the ``Brownian-like'' stochastic motion of the
core in response
to the convective ``boiling'' of the mantle of the protoneutron star during the
post-bounce, pre-explosion accretion phase, and 2) the asymmetrical bounce and
explosion
of an aspherically collapsing Chandrasekhar core.
In principle,
either phenomenon can leave the young neutron star with a speed of hundreds of
kilometers
per second.  However, neither has yet been adequately simulated or explored.
The two-dimensional radiation/hydrodynamic calculations we present here provide
only
crude estimates of the potential impulses due to mass motions and neutrino
emissions.
A comprehensive and credible investigation will require fully three-dimensional
numerical simulations not yet possible.  Nevertheless, we have in the
asymmetric
hydrodynamics of supernovae a natural means of imparting respectable kicks to
neutron stars
at birth, though speeds approaching 1000 km/s are still problematic.
\end{abstract}

\section*{Introduction}
Recent data on pulsar proper motions \cite{hla93}, a recalibration of the
pulsar distance scale \cite{tc93}, and a general recognition that previous
pulsar
surveys were biased towards low speeds \cite{ll94} imply that many pulsars have
high velocities.  Mean three-dimensional galactic speeds of 450$\pm$90 km/s
\cite{ll94}
have been estimated, with measured transverse speeds of individual pulsars
ranging from
zero to as much as $\sim$1500 km/s.  Impulsive mass loss in spherically
symmetric supernova explosions in
binaries has long been known to impart to the remaining neutron stars
velocities that reflect orbital speeds
\cite{ggo70,rs85}.  The magnitude of these kicks depends upon the
characteristics of the
binaries at explosion, which in turn depend upon mass transfer and loss and
common
envelope evolution before explosion.  If the explosion is spherical, given the
binary masses, the
orbital eccentricity, and the separation at explosion, the speed of the young
neutron
star with respect to the system center of mass is uniquely determined.

However, Dewey and Cordes \cite{dc87} performed
Monte Carlo simulations of the evolution of progenitor binaries under the
assumption that the explosions were
spherical and
were not able to reproduce the observed distribution of pulsar scintillation
velocities \cite{c86}.
They were compelled to posit an intrinsic ``kick'' with a mean magnitude of
$\sim$100 km/s, in addition
to the speed naturally imparted due to a binary origin.
Such calculations depend sensitively upon the assumed mass function, mass ratio
distribution,
initial semi-major axis and eccentricity distributions, and mass loss and
common-envelope algorithms.
Recently, Iben \& Tutukov \cite{it95} have constructed a ``scenario model''
with
state-of-the-art inputs and conclude that an ``ad hoc universal kick'' is not
required.  They
claim to reconcile the
presence of neutron stars in globular clusters and wide binaries and the
statistics of OB runaway stars, HMXB's, LMXB's, and radio pulsars
with spherical explosions.  However,
the pulsar speed distribution they derive has a mean near 100--150 km/s,
marginally consistent only
with the old pulsar distance scale \cite{tc93}, ignoring the bias of pulsar
searches to the plane \cite{ll94}.
They are able to produce maximum speeds near 1000 km/s from the explosion of a
$\sim$16 $M_\odot$
helium core in a tight binary, but are unable to reproduce the large fraction
of pulsars with speeds
above 500 km/s now inferred (reference 3 and J. Cordes, private communication).

There has long been indirect evidence that neutron stars are given a kick at
birth.  Burrows \& Woosley
\cite{bw86} concluded that if the progenitor system of PSR B1913+16 underwent
common envelope evolution
and the pre-explosion orbital eccentricity was zero, the current observed
neutron star masses and orbit
parameters are inconsistent with both a spherical explosion and known helium
core radii.  The pre-explosion
Roche limit would have been well within the progenitor helium core before the
explosion and spiral-in
would have been unavoidable.  They concluded that an intrinsic kick of at least
170 km/s
was indicated.  The same analysis with the same conclusions can be done for PSR
B1534+12. (However,
if there were no common envelope phase and the pre-explosion eccentricity were
not zero, an intrinsic
kick would not be required \cite{bw95}.)  Recently, Wasserman, Cordes, and
Chernoff (this conference) have
shown that in PSR B1913+16 the observed misalignment of the pulsar spin axis
with the orbit axis is best
explained if the pulsar received an intrinsic kick of as much as $\sim$800
km/s.  Furthermore, van den Heuvel \& Rappaport
\cite{vr86} have argued that the large eccentricities of Be/Neutron star
binaries imply the existence of
intrinsic kicks.

That most supernova remnants did not seem to have neutron stars, neither radio
pulsars nor point X-ray
sources, within them has long encouraged speculation concerning the yield of
black holes in supernova
explosions \cite{hb84,bb94}.  However, Caraveo \cite{c93} and Frail, Goss \&
Whiteoak \cite{fgw94}
have recently identified pulsars with young SNR's in a majority of SNR's with
putative ages less than 20,000
years.  These identifications depend only on high relative transverse
velocities with an inferred average value
of $\sim$500 km/s.  Excitement with these new
associations must be tempered by at least three caveats.  First, the actual
proper motions of these pulsars
have yet to be measured.   Second, the centroid of the SNR reflects in part the
mass distribution of the ISM
into which the supernova exploded and may not coincide with the position of the
actual explosion.  Third,
when a supernova explodes in a binary, the kick due to orbital motion is in the
opposite direction
to the recoil of the ejecta.  The inferred pulsar transverse speed is actually
the relative speed between the
debris and the pulsar which could be a factor of one and a half or two times
larger than the pulsar's speed
with respect to the center of mass.  Therefore, the large inferred speed may
not indicate a large intrinsic
kick, but a modest orbital kick in a spherical explosion.

Related to the emerging SNR/pulsar associations are the new data on the
``systemic'' velocities of young
supernova remnants \cite{k89,m95} (with ages from only hundreds to a few
thousand years).  It is
observed that the ``center of mass'' velocity of oxygen clumps in these
explosions is different from that of the local
ISM by -500 km/s, +900 km/s, +370 km/s, and -500 km/s for N132D, Cas A,
SN0540-69.3, and E0102.2-7219, respectively \cite{m95}.
These data imply that the explosion itself was asymmetric, but asymmetries in
the ISM and a high progenitor
speed with respect to the ISM can not yet be ruled out.  It is intriguing to
speculate that the vector
velocity of the neutron star or black hole residue of these explosions could be
(anti)correlated with
such systemic velocities.  Nevertheless, there are many other indications that
supernova explosions are
aspherical and asymmetrical.  Utrobin {\it et al.\ } \cite{u94} interpret the
``Bochum'' event in H$\alpha$ in
SN1987A with a 10$^{-3}$ M$_\odot$ shard of $^{56}$Ni moving at $\sim$4700
km/s.  The jagged optical and IR
line profiles of SN1987A, the intrinsic polarization of spectral features in
SN1993J and SN1987A, the oblateness
of recent HST images of SN1987A, and the shrapnel observed in the Vela SNR
\cite{a95}, all hint that the
explosions are asymmetrical.  Such asymmetries may have counterparts in
correlated neutron star recoils.

\section*{Previous Theories for Intrinsic Pulsar Kicks}

Theories concerning the origin of intrinsic neutron star kicks are few and
generally undeveloped.  They can
be divided into those that rely primarily on some aspect of the neutron star's
magnetic field and those
related to the supernova's dynamics or neutrino
emissions.  The former class are in part motivated by the suggestion that
pulsar proper motions are correlated
with their measured magnetic dipole moments \cite{las82}.  Data in support of
this hypothesis are no
longer compelling, particularly when selection biases are acknowledged
\cite{hla93}.  Nevertheless, the proposed
magnetic models have been clever and are worthy of review.  Harrison \&
Tademaru \cite{ht75} suggested that
the magnetic dipole of a neutron star could be off-center.  If the star
rotated, not only would the canonical
magnetic dipole radiation be emitted, but a net linear momentum would be
radiated.  The neutron star would recoil
and it could be accelerated for hundreds of years.  This theory predicted a
correlation between the pulsar's
spin axis and its proper motion that some do not find \cite{las82} and relies
on short birth spin periods to
achieve high speeds.

Chugai \cite{c84} noted that if the neutrino had a magnetic moment (and, hence,
a mass), the interaction of the
emerging neutrinos with a strong young pulsar field could lead to anisotropic
neutrino emission that could
accelerate the neutron star.  However, fields above 10$^{14}$ gauss were
required to provide a useful kick and it
is known that the dipole fields of the fastest pulsars are ``unexceptional.''
A variation on this model is provided by
Bisnovatnyi-Kogan (this volume), who relies upon the strong and anisotropic
magnetic field to alter the opacity
of the protoneutron star matter from which the emitted neutrinos decouple.

Woosley \cite{w87} published that if the integrated emission of
$\sim$3$\times$10$^{53}$ ergs in neutrinos were
radiated with an anisotropy of but 1\%, recoil speeds of $\sim$300 km/s were
possible.  This simple fact highlights the great
potential the 0.15 M$_\odot$ of relativistic neutrinos has to accelerate the
nascent neutron star.  However, these
neutrinos are radiated over many seconds and it is not clear what processes can
maintain an emission asymmetry such
that a {\it net} asymmetry of percents survives.  Note that during the first
seconds in the life of the
protoneutron star, only about one half to one third of the neutrino energy is
radiated.  It may be during this
early phase that ``convective'' motions can produce neutrino hot spots, but
these hot spots wander and the
magnitude of the net emission asymmetries during early vigorous convection is
unclear.  We have recently calculated
this effect, but not definitively, and we present our results in the next
section.  It could be that rapid rotation
plus convective overturn interact to give respectable neutrino recoils.
Shimizu, Yamada, \& Sato \cite{ss94} have recently
estimated the magnitude of rotation-induced neutrino emission anisotropies, but
too much concerning this effect
remains unresolved to draw clear conclusions.

Instabilities in the protoneutron star prior to and during the supernova
explosion have recently been implicated in
supernova explosions \cite{bhf95,hbhfc94,jm95}.  This led Herant, Benz, \&
Colgate \cite{hbc92} to posit a cascade
to an ``$\ell$ = 1'' mode as the supernova develops.  Such a dipole mode in the
hydrodynamics would imply a neutron star
recoil.  However, Burrows, Hayes, \& Fryxell \cite{bhf95} see no such cascade
to lower-order modes, and Khokhlov \cite{k91}
has identified such a cascade as an artifact of 2-D hydrodynamics.

\section*{Recoils due to Hydrodynamic Motions and Neutrino Anisotropies}

The velocity a neutron star ends up with is a vector sum of many contributions:
the galactic rotation, the
progenitor's motion with respect to the galactic rotation, the orbital kick (or
kicks, if
there were two supernovae in the binary), the deceleration due to motion in the
galactic potential, and any intrinsic
kicks imparted.  It is the perceived necessity of intrinsic kicks that
motivates this work.

Spherical explosions do not kick the residual neutron star.  However,
asymmetries in the collapse and before and after the reignition of the
supernova have the potential to impart to the core respectable recoils via
either mass motions
or anisotropic neutrino emissions.  The latter could be a consequence
of anisotropic radial opacity profiles or anisotropic accretion luminosities.
That core-collapse supernovae are subject
to Rayleigh-Taylor instabilities has been independently demonstrated by at
least three theoretical groups \cite{bhf95,hbhfc94,jm95}.
The pre-explosive core experiences a convective ``boiling''  phase behind the
temporarily stalled shock and the explosion,
when it occurs, erupts in bubbles, fingers, and plumes.  To date, these
hydrodynamic calculations have been performed
in only two dimensions (with axial symmetry) and with a variety of simplifying
approximations.  Only three-dimensional
calculations with much-improved neutrino transfer and realistic 3-D progenitor
cores that incorporate rotation and
the hydrodynamics of pre-collapse convective burning can adequately address the
issue of the kick imparted to the
residue.  Nevertheless, this class of ``simpler'' 2-D simulations can help
theorists explore the potential role of
hydrodynamics and neutrinos in imparting proper motions.  The  observed pulsar
speeds of $\sim$500 km/s are only a few
percent of the speeds ($\sim$30,000 km/s) achieved during the pre-explosive
convective phase and during the explosive phase.
In this context, the observed high pulsar speeds seem low and possible.

The {\bf star} calculation highlighted in Burrows, Hayes, \& Fryxell
\cite{bhf95} simulates in two dimensions the spherical collapse
of the core of a 15 M$_\odot$ progenitor \cite{ww95}, the early ``mantle''
convection and accretion phase
(during which the shock pauses), and the subseqent explosion.  Asymmetries,
Rayleigh-Taylor instabilities, and
overturning motions abound in this protoneutron star.  The character of this
dynamics is described in \cite{bhf95}
and in Burrows \& Hayes \cite{bh95} and will not be detailed here.  This
calculation is not best-suited to explore
recoils, since it is in 2-D, the central core ($<$ 15 km) is fixed and is
treated in 1-D, and only a 90$^{\circ}$ wedge
(with circular symmetry) from 45$^{\circ}$ to 135$^{\circ}$ in the spherical
coordinate, $\theta$, is followed
(the calculation is ``capless.'').  Nevertheless, around the core the accretion
rate per area is fluctuating
wildly and the pressure field is aspherical.  One can calculate the total
momentum of all the matter exterior to the
core and, using momentum conservation, infer the impulse to the core.  Note
that since periodic boundary conditions
are used in the angular direction and the core is fixed (and, hence, can absorb
momentum), the momentum of the outer
material from 15 to 4500 kilometers need not be zero. (This is a subtle point.)
Shown in Figure 1 is the inferred core
recoil velocity in the polar (z) direction due only to the wedge's fluctuating
motions exterior to the core.  The
corresponding value in the ``x'' direction is zero due to azimuthal symmetry.

\begin{figure} 
\vspace*{3.50in}
\hbox to\hsize{\hfill\includegraphics{vkick.ps}\kern-3in\hfill}
\caption{ The inferred velocity (in km/s) in the polar (z) direction with
which the core recoils in response to the overturning motions in the {\bf star}
calculation of Burrows, Hayes, \& Fryxell \protect\cite{bhf95}.
Note that the cap from $\theta$ equals
0$^{\circ}$ to 45$^{\circ}$ is missing in this calculations. }
\end{figure}

Figure 1 depicts two components, one that fluctuates on overturn timescales of
3--5 milliseconds, and a longer-term component that
first grows to 100--180 km/s and then subsides to near zero.  The timescale of
the latter is the supernova delay timescale of
$\sim$100 milliseconds to peak.  The fluctuating component represents the
``Brownian'' shaking that the core must
experience during the pre-explosive boiling and early explosion phases.  The
secular term grows, then shrinks because the impulse
to the core is the product of a ``mass'' with a ``velocity.'' The mass between
the shock and the core's neutrinosphere
steadily decreases with time during the delay phase, while the vigor (speed) of
the overturning and fluctuating motions
is steadily increasing.  The product peaks and then decreases.  In this
simulation, the explosion happened near t = 309
milliseconds (just after peak).  (In reality, the core's recoil need not
subside much nor asymptote to zero, but
should decouple in a way that it can not in this calculation because of the
constraints.)  While the {\bf star} calculation
is inadequate to truly explore neutron star recoils, it suggests that 3-D
speeds of hundreds of kilometers per second might
in fact be imparted during the pre-supernova and supernova phases by
hydrodynamic motions.  The impulse due to asymmetric
neutrino radiation was included in this calculation and by its end (t = 410
milliseconds) the neutrino
contribution to the inferred recoil was about $\sim$20\%.  It is interesting to
point out that the magnitude of the peak in
Figure 1 depends on the product of the ``convecting'' mass and the overturn
speeds (and the delay to explosion), which in turn
depend upon the density structure of the progenitor.  A denser core from a more
massive ZAMS star may achieve a higher peak recoil in the sense
described above (ceteris paribus) and there may be an interesting relationship
between a pulsar's proper motion and its
progenitor mass.  More realistic calculations are clearly needed to investigate
this \cite{jm94}.

The {\bf star} calculation of Figure 1 was done assuming that the collapsing
``Chandrasekhar'' core and implosion
were spherical.  Recent hydrodynamic calculations \cite{ba94} of convection
during the shell oxygen and silicon burning
that immediately precedes core collapse in massive stars suggest that
asymmetries at collapse in density, velocity, and
composition can be larger than a simple mixing-length prescription would imply.
 Furthermore, rotation might interact
with convection to further distort the core \cite{cl94}.  The upshot might be
an asymmetrical, aspherical collapse.  If the
amplitudes of the asymmetries in density or velocity are large ($\sim$percents)
and if a significant low-order mode ($\ell=1$)
exists at the onset of collapse, the consequences for a pulsar's recoil can be
significant.  To explore this hypothesis, in
January of 1995 we conducted a ``toy''
calculation of aspherical collapse.  In this {\bf kick} simulation, we
artificially decreased by 15\% the density of the
Chandrasekhar core exterior to 0.9 M$_\odot$ and within 20$^{\circ}$ of the
pole.  This calculation
was done in 2-D with azimuthal symmetry, but $\theta$ ranged between
0$^{\circ}$ and 180$^{\circ}$.  Hence, the entire core,
not just a wedge, was followed.  A 15\% decrease in density is larger than yet
seen in the calculations of Bazan \& Arnett \cite{ba94},
but we imposed no initial aspherical perturbation in velocity, despite the up
to Mach 0.25 asymmetries they have derived.
The essential point is that initial asphericities in the {\bf kick} run grew
during collapse, so that the mass column depths in
various angular directions diverged.  The matter collapsed at different rates
in different directions, though pressure forces
were transmitted in the angular directions as well that partially smoothed the
deviations.  Figure 2 depicts the core
early in the collapse.  The bounce was delayed on the side of the perturbation
wedge and the resulting shock bowed out
in the wedge direction.  The accretion rates through this shock were highly
aspherical.  To avoid burning CPU
in what was merely a toy ``proof-of-principle'' calculation, we artifically
hardened the emergent neutrino spectrum
to facilitate an early explosion.  The electron-type spectra were assumed to
have ``$\eta$'s'' of 3, above the 1.5--2.0 normally
encountered in fits to more realistic spectra \cite{mb90}.  Since neutrino
heating drives supernovae, this ignited the
explosion within 10 milliseconds of bounce.  The subsequent explosion was
aspherical not only due to the normal instabilities,
but also due to the asphericity of the matter into which the explosion emerged
and/or was driven.

\begin{figure} 
\vspace*{4.50in}
\hbox to\hsize{\hfill\includegraphics{enty3.ps}\kern-3in\hfill}
\caption{A grey-scale rendition of the entropy distribution early in the
collapse of the core constructed
for the {\bf kick} simulation.  Velocity vectors are superposed. Note the wedge
cut out near the pole at the left.
The physical scale is 2000 km from the center to the edge.
Darker color indicates lower entropy.}
\end{figure}

\begin{figure} 
\vspace*{6.00in}
\hbox to\hsize{\hfill\includegraphics{catscan2.ps}\kern+5in\hfill}
\caption{A grey-scale rendering of the entropy distribution at the end of the
{\bf kick} simulation, about 50
milliseconds into the explosion. Note the pronounced left-right asymmetry in
the ejecta and the velocity
field (as depicted with the velocity vectors).  The lengths of the velocity
vectors in Figures 2 and 3 have different scalings.
The physical scale is 2000 km from the center to the edge.
Darker color indicates lower entropy and the $\theta = 0$ axis points to the
left.}
\end{figure}

Figure 3 depicts the flow late in the
explosion.  The explosion erupted preferentially through the path of least
resistance, {\it i.e.} in the direction of the
wedge that we had imposed.  In the {\bf kick} simulation, this wedge collapsed
more slowly than the rest of the core.
Since neutrino heating drives the explosion, matter heated near the
neutrinosphere expanded
out as if from a reaction chamber.  The protoneutron star residue received a
significant impulse \`a la the rocket effect.
Furthermore, the core bounced asymmetrically and even without the neutrinos the
central residue would have recoiled away from the
direction of retarded collapse.
(It is possible that the purely hydrodynamic effect is larger than the rocket
effect, but we have yet to explore either satisfactorily.)
The impulse on the core versus time and the inferred recoil speed are depicted
in Figures 4 and 5.  An
asymmetry in collapse translates into a clear kick, though the initial total
momentum be almost zero.
Initially, the recoil is in the direction of the wedge since the rest of the
matter bounced first.  Afterwards, as the
shocked matter starts to squirt through the region of least resistance and
column depth, the recoil changes sign
and grows inexorably to its asymptotic value.  Figure
4 shows that the recoil speed in the {\bf kick} simulation reached
$\sim$500--600 km/s.  This is large, but only 2\% of the
speed of the supernova ejecta.  The contribution of neutrino radiation
asymmetry to this kick is also depicted in Figures
4 and 5 and amounts to $\sim$20\%.  It is in the same direction as the mass
motion effect, due to the fact that fluxes
are larger on the thin, or low-column, side.  Whether this is generically true
remains to be seen.

\begin{figure} 
\vspace*{3.5in}
\hbox to\hsize{\hfill\includegraphics{impulse.ps}\kern-3in\hfill}
\caption{The impulse (in{\it $\,$cgs}) imparted to the core versus time (in
seconds) in the {\bf kick} simulation.  The initial momentum is
approximately zero, but grows systematically after bounce in the direction
opposite to the artificial wedge, cut into the
core to mimic an asymmetry just before collapse. Shown are the impulses due to
neutrino asymmetry (dashed), mass motions
(dotted), and the sum (solid).}
\end{figure}

The major conclusion of the toy {\bf kick} simulation is that initial collapse
and core asymmetry can translate into
an appreciable neutron star recoil due to the variation in the collapse time
with angle, the asymmetrical bounce, the
variation in the tamp with angle, and the rocket effect.  Quite naturally,
the debris and the residue
move in opposite directions, conserving total momentum.  Since
accretion-induced
collapse is not preceded by the convective burning stages characteristic of the
final hours of the core of a massive star, the initial
asymmetries in the two contexts may be quite different.  Consequently, the
proper motions of AIC neutron stars and those of
neutron stars from massive stars may be systematically different, with those of
the latter being on average higher.  However, to explore
these phenomena in more detail and more credibly will require a realistically
aspherical initial core, 3-D simulations, and
better neutrino transfer.  Nevertheless, it is heartening that proper motions
of the ``correct'' magnitude are produced in
these embryonic simulations of multi-dimensional supernovae.

Given the stochastic nature of the processes we have highlighted here by which
we have attempted to explain intrinsic pulsar
kicks, we expect that Nature provides not a single high kick speed, but a broad
distribution of speeds.  These will depend
upon the degree and character of the initial asymmetries, the initial rotation
structure, the duration of the delay to
explosion, the progenitor density profiles, and chance.

\begin{figure} 
\vspace*{3.5in}
\hbox to\hsize{\hfill\includegraphics{recoil.ps}\kern-3in\hfill}
\caption{Same as Figure 4, but the inferred recoil speed (in km/s) of the
residual neutron star versus time (in seconds).}
\end{figure}

\acknowledgments
The authors would like to thank T. Janka, W. Benz, I. Iben, and G. Bazan for
stimulating conversations
during the germination of this work and the N.S.F. for financial support under
grant \# AST92-17322.


\begin{references}
\bibitem{hla93}P.A. Harrison, A.G. Lyne, \& B. Anderson, Mon. Not. R. Astron.
Soc. {\bf 261}, 113 (1993).
\bibitem{tc93}J.H. Taylor \& J.M. Cordes, Astrophys.\ J., {\bf 411}, 674
(1993).
\bibitem{ll94}A. Lyne \& D.R. Lorimer, Nature {\bf 369}, 127 (1994).
\bibitem{ggo70}J.R. Gott, J.E. Gunn, \& J.P. Ostriker, Astrophys.\ J. {\bf
160}, L91 (1970).
\bibitem{rs85}V. Radhakrishnan \& C.S. Shukre, in {\it Supernovae, Progenitors,
and Remnants}, eds.
G. Srinivasan \& V. Radhakrishnan (Indian Academy of Sciences: Bangalore),
pp.155--169 (1985).
\bibitem{dc87}R.J. Dewey \& J.M. Cordes, Astrophys.\ J. {\bf 321}, 780 (1987).
\bibitem{c86}J.M. Cordes, Astrophys.\ J. {\bf 311}, 183 (1986).
\bibitem{it95}I. Iben \& A. Tutukov, submitted to Astrophys.\ J., (1995).
\bibitem{bw86}A. Burrows \& S.E. Woosley, Astrophys.\ J. {\bf 308}, 680 (1986).
\bibitem{bw95}G.E. Brown \& J.C. Weingartner, Astrophys.\ J. {\bf 436}, 843
(1995).
\bibitem{vr86}E.P.J. van den Heuvel \& S. Rappaport, in {\it I.A.U. Colloquium
92}, eds. A. Slettebak \& T.D. Snow
(Cambridge Univ. Press), pp. XXX (1986).
\bibitem{hb84} D.J. Helfand \& R.H. Becker, Nature {\bf 307}, 215 (1984).
\bibitem{bb94}G.E. Brown \& H.A. Bethe, Astrophys.\ J. {\bf 423}, 659 (1994).
\bibitem{c93}P. Caraveo, Astrophys.\ J. {\bf 415}, L111 (1993).
\bibitem{fgw94}D.A. Frail, W.M. Goss, \& J.B.Z. Whiteoak, Astrophys.\ J. {\bf
437}, 781 (1994).
\bibitem{k89}R.P. Kirchner, J.A. Morse, P.F. Winkler, \& J.P. Blair,
Astrophys.\ J. {\bf 342}, 260 (1989).
\bibitem{m95}J.A. Morse, P.F. Winkler, \& R.P. Kirchner, submitted to
Astrophys.\ J. (1995).
\bibitem{u94}V.P. Utrobin, N.N. Chugai, \& A.A. Andronova, submitted to Astron.
\& Astrophys. (1994).
\bibitem{a95} B. Aschenbach, R.J. Egger, \& J. Trumper, Nature {\bf 373}, 587
(1995).
\bibitem{las82}A.G. Lyne, B. Anderson, \& M.J. Salter, Mon. Not. R. Astron.
Soc. {\bf 201}, 503 (1982).
\bibitem{ht75}E.R. Harrison \& E. Tademaru, Astrophys.\ J. {\bf 201}, 447
(1975).
\bibitem{c84}N.N. Chugai, Sov. Astron. Lett. {\bf 10}, 87 (1984).
\bibitem{w87}S.E. Woosley, in {\it The Origin and Evolution of Neutron Stars},
eds. D.J. Helfand \& J.-H. Huang
(D. Reidel: Dordrecht), p. 255 (1987).
\bibitem{ss94}T. Shimizu, S. Yamada, \& K. Sato, Astrophys.\ J. {\bf 434}, 268
(1994).
\bibitem{bhf95}A. Burrows, J. Hayes, \& B.A. Fryxell, Astrophys.\ J., in press
(1995).
\bibitem{hbhfc94}M. Herant, W. Benz, J. Hix, C. Fryer, \& S.A. Colgate,
Astrophys,\ J. {\bf 435}, 339 (1994).
\bibitem{jm95}H.-T. Janka \& E. M\"uller, Astron. \& Astrophys., in press
(1995).
\bibitem{hbc92}M. Herant, W. Benz, \& S.A. Colgate, Astrophys.\ J. {\bf 395},
642 (1992).
\bibitem{k91}A. Khokhlov, Astron. \& Astrophys. {\bf 290}, 496 (1991).
\bibitem{ww95}T. Weaver \& S.E. Woosley, Astrophys.\ J. Suppl., in press
(1995).
\bibitem{bh95}A. Burrows \& J. Hayes, to appear in the
proceedings of the {\it 17'th Texas Symposium on Relativistic Astrophysics},
held in Munich, Germany,
December 12-16, 1994 in the
Annals of the New York Academy of Sciences,
ed. H. Boeringer.
\bibitem{jm94}H.-T. Janka \& E. M\"uller, Astron. \& Astrophys. {\bf  },
(1994).
\bibitem{ba94}G. Bazan \& D. Arnett, Astrophys.\ J. {\bf 433}, 41 (1994).
\bibitem{cl94}S.A. Colgate \& P.J.T. Leonard, in {\it Gamma-Ray Bursts}, eds.
G.J. Fishman, J.J. Brainerd,
\& K. Hurley (A.I.P. Press), p. 581 (1994).
\bibitem{mb90}E. Myra \& A. Burrows, Astrophys.\ J. {\bf 364}, 222 (1990).


\end{references}
\end{document}